\documentclass{article}
\usepackage{spconf}
\usepackage{amsmath,amsfonts,amsthm,amssymb}
\usepackage{graphicx,float,wrapfig}
\usepackage{graphicx}
\usepackage{graphics,color}
\usepackage{mathtools}
\usepackage{nicefrac}
\usepackage[T1]{fontenc}
\usepackage{upquote}
\usepackage{algorithm,algorithmic}
\usepackage{caption}% http://ctan.org/pkg/caption
\makeatletter
\@addtoreset{algorithm}{section}% algorithm counter resets every chapter
\makeatother
\newtheorem{theorem}{Theorem}[section]
\newtheorem{definition}[theorem]{Definition}

\numberwithin{equation}{section}

\def\x{{\mathbf x}}

\title{Precise Semidefinite Programming Formulation of Atomic Norm Minimization for Recovering d-Dimensional ($d\geq 2$) Off-the-Grid Frequencies}
\name{Weiyu Xu, Jian-Feng Cai, Kumar Vijay Mishra, Myung Cho, and Anton Kruger}
\address{The University of Iowa, Iowa City IA 52242}

\begin{document}
\ninept
\maketitle
\begin{abstract}
Recent research in off-the-grid compressed sensing (CS) has demonstrated that, under certain conditions, one can successfully recover a spectrally sparse signal from a few time-domain samples even though the dictionary is continuous. In particular, atomic norm minimization was proposed in  \cite{tang2012csotg} to recover $1$-dimensional spectrally sparse signal. However, in spite of existing research efforts \cite{chi2013compressive}, it was still an open problem how to formulate an equivalent positive semidefinite program for atomic norm minimization in recovering signals with $d$-dimensional ($d\geq 2$) off-the-grid frequencies. In this paper, we settle this problem by proposing equivalent semidefinite programming formulations of atomic norm minimization to recover signals with $d$-dimensional ($d\geq 2$) off-the-grid frequencies.
\end{abstract}
\section{Introduction}
\label{sec:intro}
Compressed sensing (CS) is a new sampling paradigm which promises to unite two critical steps involved in processing signals: digital data acquisition and its compression \cite{donoho2006compressed} \cite{kutyniok2012theory}. To recover a signal from fewer random measurements, CS algorithms harness the inherent sparsity of the signal under some appropriate basis or dictionary.

We consider $d$-dimensional signal,where $d\geq 2$. For example, consider a frequency-sparse signal $x^{\clubsuit}[\mathbf{l}]$ represented as a sum of $s$ complex exponentials,
\begin{equation}
\label{eq:sigmodelstd}
x^{\clubsuit}[\mathbf{l}] = \sum\limits_{j=1}^{s} c_je^{i2\pi \mathbf{f}_j^T\mathbf{l}} = \sum\limits_{j=1}^{s} |c_j|a(\mathbf{f}_j, \phi_j)[\mathbf{l}]\phantom{1}, \phantom{1} \mathbf{l} \in \mathcal{N}
\end{equation}
where $d$ is the signal dimension,  $c_j = |c_j|e^{i\phi_j}$ ($i = \sqrt{-1}$) represents the complex coefficient of the frequency $\mathbf{f}_j \in [0, 1]^d$, (with amplitude $|c_j| > 0$ and phase $\phi_j \in [0, 2\pi)$), and frequency-\textit{atom} $a(\mathbf{f}_j, \phi_j)[\mathbf{l}] = e^{i(2\pi \mathbf{f}_j^T \mathbf{l} + \phi_j)}$. We use the index set $\mathcal{N} = \prod_{p=1}^{d} \{l\phantom{1}|\phantom{1} 0 \le l \le n_p-1\}$ to represent the set of time indices of the signal, where $n_p \in \mathbb{N}$ for $1\leq p \leq n$, and $|\mathcal{N}| = \prod_{p=1}^{d} n_p$. Generally, we have $n_p=n$, for every $1\leq p\leq d$. It is customary to label only the frequency information - either the exponentials $e^{i2\pi \mathbf{f}_j^T\mathbf{\mathbf{l}}}$ or just $\mathbf{\mathbf{f}}_j$ - as $poles$ \cite{majda1989simple} \cite{wei1990new}.

When $\mathbf{f}_j$ takes values only on a \textit{discrete frequency grid}, the Discrete Fourier Transform (DFT) matrix can be used as an appropriate finite discrete dictionary for the sparse representation of $x[\mathbf{l}]$. However, it is quite possible for the true frequencies to be anywhere in the \textit{continuous domain} $[0, 1]$. Since the true continuous-domain frequencies may lie off the center of the DFT bins, the DFT representation in this case would destroy the sparsity of the signal and result in the so-called \textquotedblleft basis mismatch\textquotedblright\phantom{1}\cite{chi2011sensitivity}. Traditionally it was believed that finer discretization of the DFT grid would not get rid of this basis mismatch problem, by leading to higher correlation of the sensing matrix and, thus, computationally infeasible or expensive signal recovery \cite{chi2011sensitivity, duarte2013spectral}. Nevertheless, state-of-the-art results have shown that one can indeed effectively tackle the problem of basis mismatch by discretizing the continuous dictionaries under very general conditions \cite{tang2013justdiscretize}.

The numerical problems associated with the spectral spill-over in the Dirichlet kernel have also recently been addressed by the \textit{off-the-grid compressed sensing} approach \cite{candes2013towards} \cite{tang2012csotg} for signals with $1$-dimensional off-the-grid frequency. This method relies on atomic norm minimization and guarantees recovery of frequencies lying anywhere in the continuous domain [0, 1] from a limited number of random observations, provided the line spectrum satisfies nominal resolution conditions. For recovering off-the-grid frequencies in $2$-dimensional signals, \cite{chi2013robust} proposed to use Hankel matrix completion which guarantees robustness against corruption of data. We remark that the method used in \cite{chi2013robust} is not atomic norm minimization. In \cite{chi2013compressive}, the authors studied the theoretical performance of hypothetical atomic norm minimization for signals with $2$-dimensional off-the-grid frequencies, and proposed a \emph{heuristic} semidefinite program to \emph{approximate} the atomic norm minimization computationally. However, the heuristic semidefinite programming in \cite{chi2013compressive} is not guaranteed to provide the atomic norm minimization in general. As noted in \cite{chi2013compressive}, ``Unfortunately, the exact semidefinite programming characterization of atomic norm minimization for line spectrum estimation, as proposed in \cite{tang2012csotg}, cannot be extended to 2D in the most general sense. This arises due to the fundamental difficulty of generalizing the Caratheodory's theorem beyond the 1D model.''  In fact, for $1$-dimensional signal, the proof of the semidefinite programming in \cite{tang2012csotg} (see also (\ref{eq:semiotg}) in this paper) being equivalent to atomic norm minimization relies on the Vandemonde decomposition for Toeplitiz matrix by Carath{\`e}odory lemma \cite{tang2012csotg}. However, for $d$-dimensional ($d\geq 2$) signal, the Vandemonde decomposition does not extend to block Toeplitz matrices with Toeplitiz blocks. So a direct extension from Toeplitz matrices in (\ref{eq:semiotg}) to block Toeplitz matrices with Toeplitz blocks does not give the atomic norm in higher-dimensional frequencies \cite{chi2013compressive}. Furthermore, how to set up precise semidefinite programming to perform atomic norm minimization for signals with $2$-dimensional off-the-grid frequencies was not known in the prior literature, to the best of our knowledge.

In this paper, we settle this problem by proposing equivalent semidefinite programming formulations of atomic norm minimization to recover signals with $d$-dimensional ($d\geq 2$) off-the-grid frequencies. We remark that our results are applicable to an arbitrary signal dimension $d\geq 2$. Our results are also extensible to noisy observations. 

The remainder of this paper is organized as follows. In Section \ref{sec:sysmodel}, we formally introduce the system model. In Section \ref{sec:ouralg}, we introduce equivalent positive semidefinite programming to minimize atomic norm for $d$-dimensional ($d\geq 2$) signals. In Section \ref{sec:numsim}, we give numerical simulations.

\section{System Model}
\label{sec:sysmodel}
The signal in (\ref{eq:sigmodelstd}) can be modeled as a positive linear combination of the unit-norm frequency-\textit{atoms} $a(\mathbf{f}_j, \phi_j)[\mathbf{l}] \in \mathcal{A} \subset \mathbb{C}^{\mathcal{N}}$ where $\mathcal{A}$ is the set of all frequency-atoms, and $\mathbf{f}_j=(\mathbf{f}_{j,1}, \mathbf{f}_{j,2},...,\mathbf{f}_{j,d})^T \in [0,1]^d$. These frequency atoms are basic units for synthesizing the frequency-sparse signal. Further, suppose the signal in (\ref{eq:sigmodelstd}) is observed on the index set $\mathcal{M} \subset \mathcal{N}$, $|\mathcal{M}| = m \ll \prod_{p=1}^{d}n_p$ where $m$ observations are chosen uniformly at random. Then, for $d=1$-dimensional signal, to estimate the remaining samples of the signal $x$ over $\mathcal{N} \setminus \mathcal{M}$, \cite{chandrasekaran2012theconvex, tang2012csotg} suggests minimizing the atomic norm $||\hat{x}||_\mathcal{A}$ - a sparsity-enforcing analog of $\ell_1$ norm for a general atomic set $\mathcal{A}$-among all vectors $\hat{x}$ leading to the same observed samples as $\x$. The atomic norm is given by,
\begin{equation}
\label{eq:atomicnorm}
||\hat{x}||_{\mathcal{A}} = \underset{c_j, \mathbf{f}_j}{\text{inf}}\phantom{1}\left\{\sum\limits_{j=1}^s|c_j|: \hat{x}[\mathbf{l}] = \sum\limits_{j=1}^{s} c_je^{i2\pi \mathbf{f}_j^T\mathbf{l}} \phantom{1}, \phantom{1} \mathbf{l} \in \mathcal{M}\right\}
\end{equation}
For $1$-dimensional signal, the semidefinite formulation of $||\hat{x}||_\mathcal{A}$ is defined as follows:
\begin{definition} (\cite{tang2012csotg}, for 1-dimensional signal) Let $T_n$ denote the $n \times n$ positive semidefinite Toeplitz matrix, $t \in \mathbb{R}^+$, Tr($\cdot$) denote the trace operator and $(\cdot)^*$  denote the complex conjugate. Then,
    \begin{equation}
	||\hat{x}||_{\mathcal{A}} = \underset{T_n, t}{\text{inf}} \left\{\dfrac{1}{2|\mathcal{N}|} \text{Tr($T_n$)} + \frac{1}{2}t : \begin{bmatrix*}[r] T_n & \hat{x} \\ \hat{x}^* & t \end{bmatrix*} \succeq 0 \right\}
	\end{equation}
\end{definition}
From \cite{tang2012csotg}, the positive semidefinite Toeplitz matrix $T_n$ is related to the frequency atoms through the following result by Carath{\`e}odory \cite{cara1911uber}:
\begin{align}
T_n &= URU^*\\
\text{where } U_{l'j'} &= a(f_{j'}, \phi_{j'})[l'],\\
                    R &= \text{diag}([b_1, \cdots, b_{r'}])
\end{align}
The diagonal elements of $R$ are real and positive, and $r' = \text{rank}(T_n)$.

Consistent with this definition, \emph{for $1$-dimensional signal}, the atomic norm minimization problem for the frequency-sparse signal recovery can now be formulated in a semidefinite program (SDP) with $m$ affine equality constraints:
\begin{flalign}
	\label{eq:semiotg}
	& \underset{T_n, \hat{x}, t}{\text{minimize}}\phantom{1} \dfrac{1}{2|\mathcal{N}|} \text{Tr($T_n$)} + \frac{1}{2}t\nonumber\\
	& \text{subject to}\phantom{1} \begin{bmatrix*}[r] T_n & \hat{x} \\ \hat{x}^* & t \end{bmatrix*} \succeq 0\\
	& \hat{x}[l] = x^{\clubsuit}[l], \phantom{1} l \in \mathcal{M},\nonumber
\end{flalign}

For $1$-dimensional signal, the proof of (\ref{eq:semiotg}) being equivalent to atomic norm minimization relies on the Vandemonde decomposition for Toeplitiz matrix by Carath{\`e}odory lemma \cite{tang2012csotg}. However, for $d$-dimensional ($d\geq 2$) signal, the Vandemonde decomposition does not extend to block Toeplitz matrices with Toeplitiz blocks for high dimensional signals. Indeed,  a direct extension from Toeplitz matrices  in (\ref{eq:semiotg}) to block Toeplitz matrices with Toeplitz blocks does not give the atomic norm in higher-dimensional frequencies \cite{chi2013compressive}.

In this paper, we settle this problem by proposing equivalent semidefinite programming formulations of atomic norm minimization to recover signals with $d$-dimensional ($d\geq 2$) off-the-grid frequencies. Note that our results are widely applicable to an arbitrary signal dimension $d\geq 2$.

\section{Positive Semidefinite Programs for Atomic Norm Minimization in Recovering High-Dimensional Frequencies}
\label{sec:ouralg}

In this section, we set out to give positive semidefinite programming for atomic norm minimization for $d$-dimensional signal, where $d\geq 2$. The key idea is to look at the dual problem of atomic norm minimization, and then transform the dual problem to equivalent positive semidefinite programming by using theories for positive trignometric polynomials.

 Extending from $1$-dimensional signal case in \cite{tang2012csotg}, for two tensors $q$ and $x$, we define the inner product between them as $\langle q,x \rangle=\vec{x}^*\vec{q}$, where $\vec{q}$ and $\vec{x}$  mean the vectorization of $q$ and $x$,   and we also define the real part of the inner product as $\langle q,x \rangle_{\mathbb{R}}=\text{Re}(\vec{x}^*\vec{q})$. Then the dual norm of the atomic norm $\|\cdot\|_{\mathcal{A}}$ is given by
$$\|q\|_{\mathcal{A}}^*=\sup_{\|x\|_{\mathcal{A}}\leq 1} \langle q,x \rangle_{\mathbb{R}}=\sup_{\mathbf{f} \in [0,1]^d}|\langle q, a(\mathbf{f},0)\rangle|     .$$

The primal atomic norm minimization problem is given by
\begin{flalign}
	\label{eq:atomicminimization}
	& \underset{x}{\text{minimize}}\phantom{1}  \|x\|_{\mathcal{A}}\nonumber\\
%	& \text{subject to}\phantom{1} \begin{bmatrix*}[r] T_n & \hat{x} \\ \hat{x}^* & t \end{bmatrix*} \succeq 0\\
	& \text{subject to}\phantom{1} x[\mathbf{l}] = x^{\clubsuit}[\mathbf{l}], \phantom{1} \mathbf{l} \in \mathcal{M}
\end{flalign}

Similar to the derivation in \cite{tang2012csotg}, its dual problem is given by
\begin{flalign}
	\label{eq:dual toatomicminimization}
	 \underset{q}{\text{maximize}} & \phantom{1} \langle q_{\mathcal{M}},x^{\clubsuit}_{\mathcal{M}}\rangle_{\mathbb{R}}\phantom{1}  \nonumber\\
	 \text{subject to}&\phantom{1}\|q\|_{\mathcal{A}}^* \leq 1 \\
	&\phantom{1} q_{\mathcal{N}\setminus \mathcal{M}}=0 \nonumber
\end{flalign}

By the Slater's condition, strong duality holds between the primal problem and dual problem. So we do not loosen the optimal solution by turning to its dual problem.

We note that $$ \langle q, a(\mathbf{f},0) \rangle=\sum_{\mathbf{j}} q_{\mathbf{j}}e^{-i 2\pi \mathbf{f}^T \mathbf{j} },$$
where ${\mathbf{j}}=\{j_1,j_2,...,j_d\} \in \mathcal{N}$, and $0\leq j_d \leq n_p-1$ for $1\leq p \leq d$. Namely $ \langle q, a(\mathbf{f},0) \rangle$ is a $d$-variate trigonometric polynomial with variables $\mathbf{f}$. Now we are ready to describe our positive semidefinite programming to solve the dual optimization problem.

We first choose a certain sum-of-squares relaxation degree vector $\mathbf{m}=(m_1,m_2,...m_d)^T$, where $m_p \geq n_p-1$ for every $1\leq p\leq d$. We then define a zero-padded extension $\widetilde{q}$ of $q$ under $\mathbf{m}$. For $\mathbf{j}=(j_1,...,j_d) \in \prod_{p=1}^{d} \{l\phantom{1}|\phantom{1} 0 \le l \le m_p\}$, we define $\widetilde{q}_\mathbf{j}$ as follows:

\begin{equation*}
\widetilde{q}_\mathbf{j}=
\begin{cases}
q_{\mathbf{j}} &\mbox{if~} \mathbf{j} \in \mathcal{N} \cr
0 & \mbox{otherwise} .
\end{cases}
\end{equation*}

For $\widetilde{q}$, we denote its vectorization by $\vec{\widetilde{q}}$, namely for every $\mathbf{j}=(j_1,...,j_d) \in \prod_{p=1}^{d} \{l\phantom{1}|\phantom{1} 0 \le l \le m_p\}$,
$$\vec{\widetilde{q}}_{u }=\widetilde{q}_\mathbf{j},$$
where
$$u=\left(\sum\limits_{t=1}^{d}\left[ j_{d-t+1} \left(\prod \limits_{p=1}^{d-t} (m_p+1) \right) \right ] \right)+1$$
(namely $u$ is a natural number between  $1$ and $\prod_{p=1}^{d} (m_p+1)$).

We then loosely follow the notations in \cite{dumitrescu2007positive}. For each $1\leq p \leq d$, let us define $\Theta_{k_p}$ as an $(m_p+1) \times (m_p+1)$ elementary Toeplitz matrix with ones on the $k_p$-th diagonal and zeros elsewhere, where $-m_p\leq k_p \leq m_p$. We remark that we take the main diagonal of an $(m_p+1) \times (m_p+1)$ square matrix as the $0$-th diagonal, and its uppermost diagonal as the $m_p$-th diagonal.

In addition, for a $d$-tuple $\mathbf{k}=(k_1, k_2,...,k_d)$, where $-m_p\leq k_p \leq m_p$ for every $1\leq p\leq d$, we define
$$\Theta_{\mathbf{k}}=\Theta_{k_d} \otimes ... \otimes \Theta_{k_1},$$
where $\otimes$ is the Kronecker product.  

Then by the Bounded Real Lemma for multivariate trigonometric polynomials \cite{dumitrescu2007positive},
$\|q\|_{\mathcal{A}}^* < 1$ implies that for a certain degree vector $\mathbf{m}$, there exists a Hermitian matrix $Q_0 \succeq 0$ such that
$$\delta_{\mathbf{k}}=tr\left[\Theta_\mathbf{k} Q_0\right], \mathbf{k} \in \mathcal{H},$$
where $\mathcal{H}$ is a halfspace of $\prod_{p=1}^{d} \{l\phantom{1}|\phantom{1} -m_p \le l \le m_p\}$, $\delta_{\mathbf{0}}=1$, $\delta_{\mathbf{k}} =0$ if $\mathbf{k}\neq \mathbf{0}$,  and
$$\begin{bmatrix*}[r] Q_0 & \vec{\widetilde{q}} \\ \vec{\widetilde{q}}^* & 1 \end{bmatrix*} \succeq 0.$$
 It is also very easy to see that when such a positive semidefinite matrix $Q_0$ exists, $\|q\|_{\mathcal{A}}^* \leq 1$.

Then the dual problem (\ref{eq:dual toatomicminimization}) is reduced to (up to exchanging between strict and non-strict inequalities, and the choice of $\mathbf{m}$)
\begin{flalign}
	\label{eq:dual toatomicminimizationSDP}
	 \underset{q,Q_0}{\text{maximize}} \phantom{1} & \langle q_{\mathcal{M}},x^{\clubsuit}_{\mathcal{M}} \rangle_{\mathbb{R}}\phantom{1}  \nonumber\\
	 \text{subject to}\phantom{1} & \delta_{\mathbf{k}}=tr\left[\Theta_\mathbf{k} Q_0\right], \mathbf{k} \in \mathcal{H} \\
& \begin{bmatrix*}[r] Q_0 & \vec{\widetilde{q}} \\ \vec{\widetilde{q}}^* & 1 \end{bmatrix*} \succeq 0,\nonumber \\
	&     \phantom{1} q_{\mathcal{N}\setminus \mathcal{M}}=0, \nonumber
\end{flalign}
where $\widetilde{q}$ is the extension of $q$, and $\vec{\widetilde{q}}$ is the vectorization of $\widetilde{q}$ as described above.

In solving the dual problem, we need to decide the sum-of-squares relaxation degree $\mathbf{m}$.  This sum-of-squares relaxation degree then decides the dimension of the PSD matrix $Q_0$. Suppose we fix the sum-of-squares degree as $(m_1, m_2,...,m_d)^T$, then the dimension of  $Q_0$ is $(\prod_{p=1}^{d}{(m_p+1)}) \times (\prod_{p=1}^{d}{(m_p+1)})$. For $\widehat{q}$ being the optimal solution of (\ref{eq:dual toatomicminimization}),  positive trigonometric polynomial theories guarantee that there exist a finite $\mathbf{m}$ such that there exists an $(\prod_{p=1}^{d}{(m_p+1)}) \times (\prod_{p=1}^{d}{(m_p+1)})$ positive semidefinite matrix $Q_0$ satisfying the constraints of the semidefinite programming (\ref{eq:dual toatomicminimizationSDP}), (up to exchanging between strict and non-strict inequalities), and thus (\ref{eq:dual toatomicminimizationSDP}) gives the exact solution (up to exchanging strict and non-strict inequalities) to (\ref{eq:dual toatomicminimization}). We just keep increasing $\mathbf{m}$ if a lower $\mathbf{m}$ does not suffice to minimize the atomic norm in (\ref{eq:dual toatomicminimizationSDP}).

But how to decide whether a certain $\mathbf{m}$ suffices for our purpose of minimizing the atomic norm? We introduce a checking mechanism to check whether the minimum atomic norm has been achieved.  To understand this mechanism, let us call the optimization problem (\ref{eq:atomicminimization}) as the primal problem of atomic norm minimization, the optimization problem (\ref{eq:dual toatomicminimization}) as the dual problem of atomic norm minimization, and the optimization problem (\ref{eq:dual toatomicminimizationSDP}), for a certain $\mathbf{m}$, as the $\mathbf{m}$-restricted dual problem of atomic norm minimization. In fact, one can easily show that the feasible set for $q$ in (\ref{eq:dual toatomicminimizationSDP}) is a subset of that of (\ref{eq:dual toatomicminimization}).
Thus the optimal objective value $p_{\mathbf{m}\cdot{\text{dual}}}$ of (\ref{eq:dual toatomicminimizationSDP}) is no bigger than the optimal objective value $p_{\text{dual}}$ of (\ref{eq:dual toatomicminimizationSDP}), namely, 
$$ p_{\mathbf{m}\cdot{\text{dual}}}\leq p_{\text{dual}}.$$
After we get the optimal solution $q^*$ of (\ref{eq:dual toatomicminimizationSDP}), one can use the dual polynomial method \cite{tang2012csotg} (readily extended to $d \ge 2$ ) to get the $t$ frequency poles $\mathbf{f}_1,~...,~\mathbf{f}_t$ which satisfy $\langle q^*, a(\mathbf{f}_j,0) \rangle=1$, $1\leq j \leq t$.  Then we plug these $t$ frequencies back into the set of observation constraints in the primal problem:
$$x[\mathbf{l}] = x^{\clubsuit}[\mathbf{l}], \phantom{1} \mathbf{l} \in \mathcal{M},$$
to solve for the $t$ coefficients $c_j$, $1\leq j \leq t$, for these $t$ frequencies.  Then 
\begin{equation}
\label{eq:feasible}
x[\mathbf{l}] = \sum\limits_{j=1}^{t} c_je^{i2\pi \mathbf{f}_j^T\mathbf{l}} = \sum\limits_{j=1}^{t} |c_j|a(\mathbf{f}_j, \phi_j)[\mathbf{l}]\phantom{1}, \phantom{1} \mathbf{l} \in \mathcal{N}
\end{equation}
is a feasible solution to (\ref{eq:atomicminimization}), and its corresponding objective value is $p_{{\text{primal}_f}}=\sum\limits_{j=1}^{t} |c_j|$. We remark that $p_{\text{primal}_f}\geq p_{\text{primal}}$, where $p_{\text{primal}}$ is the optimal objective value for (\ref{eq:atomicminimization}). Then if $p_{{\text{primal}_f}}$ is equal to the optimal objective value $p_{\mathbf{m}\cdot{\text{dual}}}$ of (\ref{eq:dual toatomicminimizationSDP}), by weak duality,  we immediately know the $\mathbf{m}$-restricted dual problem (\ref{eq:dual toatomicminimizationSDP}) gives the minimum atomic norm.

In solving the $\mathbf{m}$-restricted dual problem, we remark that one can also solve the dual of the $\mathbf{m}$-restricted dual problem, which does not make an essential difference.

\section{Numerical Simulations}
\label{sec:numsim}
\begin{figure} \centering
\includegraphics[width=0.5\textwidth]{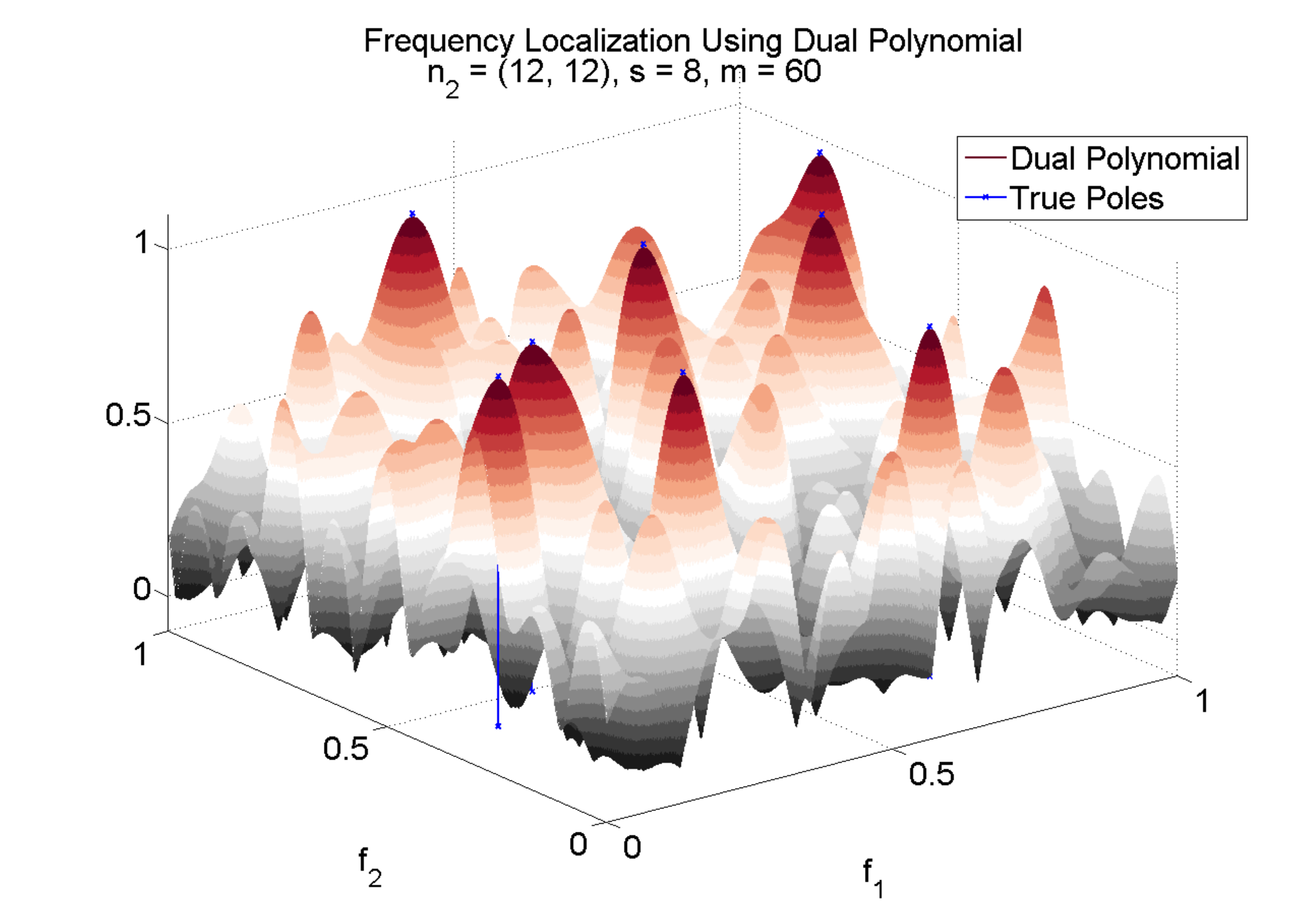}
\caption{Frequency localization for multidimensional case using dual polynomial. The dual polynomial assumes a maximum modulus of unity if $\mathbf{f}$ corresponding to the true poles.}
\label{fig:freqlocal2D}
\end{figure}

We evaluated the multi-dimensional atomic norm minimization by using SDPT3 \cite{tutuncu2003solving} to solve the semidefinite program in (\ref{eq:dual toatomicminimizationSDP}). We restricted ourselves to $d = 2$ and randomly drew $s = 8$ frequency pairs in the band $[0, 1]^2$ for the artificially generated signal. The phases of the signal frequencies were sampled uniformly at random in $[0, 2\pi)$. The amplitudes $|c_j|, j = 1, \cdots, s$ were drawn randomly from the distribution $0.5 + \chi^2_1$ where $\chi^2_1$ represents the Chi-squared distribution with 1 degree of freedom. A total of $m = 60$ observations were randomly chosen for the sample set $\mathcal{M}$ for $(n_1, n_2) = (12, 12)$. \\
Figure \ref{fig:freqlocal2D} shows the frequency localization using dual polynomial approach for this signal. We note that $ \langle q, a(\mathbf{f},0) \rangle = 1$ for $\mathbf{f}$ corresponding to the true poles. And for this example, $\mathbf{m}=(11,11)$ suffices to achieve minimizing the atomic norm, as shown by our checking mechanism devised in Section \ref{sec:ouralg}. 
%\begin{figure} \centering
%\includegraphics[width=0.5\textwidth]{freqlocalization2D.eps}
%\caption{Frequency localization for multidimensional case using dual polynomial. The dual polynomial assumes a maximum modulus of unity if $\mathbf{f}$ corresponding to the true poles.}
%\label{fig:freqlocal2D}
%\end{figure}

\section{Acknowledgement}
\label{sec:ack}
The authors would like to thank Gongguo Tang and Benjamin Recht of University of Wisconsin at Madison for helpful e-mail discussions related to their work in \cite{tang2012csotg}. Weiyu Xu would like to thank Yuejie Chi for her conference presentation at the Asilomar 2013 which introduces this open problem and inspires our curiosity about it. Weiyu Xu is also thankful to Babak Hassibi for a helpful discussion of this problem during his visit at Caltech.

\bibliographystyle{IEEEbib}
\bibliography{refs}

\end{document}